\documentclass[a4paper]{article}
\usepackage{graphicx}
\usepackage{booktabs}
\usepackage{xcolor}
\usepackage{amssymb}
\usepackage{amsfonts}

\usepackage{lineno}
\usepackage{newtxmath}
\usepackage{natbib}
\usepackage{geometry}
\usepackage{comment}
\usepackage{subfiles} 
\usepackage{academicons}
\definecolor{orcidlogocol}{HTML}{A6CE39}

\newcommand{\lapprox} {\, \lower3pt\hbox{$\sim$}\llap{\raise2pt\hbox{$<$}}\,}
\newcommand{\gapprox} {\, \lower3pt\hbox{$\sim$}\llap{\raise2pt\hbox{$>$}}\,}

\chardef\us=`\_

\begin{document}
\Large
\textbf{Variation of the electron flux spectrum along a solar flare loop as inferred from STIX hard X-ray observations}

\normalsize

Anna Volpara$^1$, Paolo Massa$^2$, S\"am Krucker$^{3,4}$, A. Gordon Emslie$^2$, Michele Piana$^{1,5}$, Anna Maria Massone$^1$ \\

\hspace{-0.5cm}$^1$ MIDA, Dipartimento di Matematica, Università di Genova, via Dodecaneso 35 16146 Genova, Italy \\
email: volpara@dima.unige.it,piana@.dima.unige.it, massone@.dima.unige.it \\
$^2$   Department of Physics \& Astronomy, Western Kentucky University, 1906 College Heights Blvd., Bowling Green, KY 42101, USA\\
email: paolo.massa@wku.edu, gordon.emslie@wku.edu \\
$^3$  University of Applied Sciences and Arts Northwestern Switzerland, Bahnhofstrasse 6, 5210 Windisch, Switzerland \\
$^4$ Space Sciences Laboratory, University of California, 7 Gauss Way, Berkeley, CA 94720, USA \\
email: krucker@berkeley.edu \\
$^5$  Istituto Nazionale di Astrofisica, Osservatorio Astrofisico di Torino, Italy \\

\date{\today}

\begin{center}
   \textbf{Abstract} 
\end{center}

   Regularized imaging spectroscopy was introduced for the construction of electron flux images at different energies from count visibilities recorded by the Reuven Ramaty High Energy Solar Spectroscopic Imager (RHESSI). In this work we seek to extend this approach to data from the Spectrometer/Telescope for Imaging X-rays (STIX) on-board the Solar Orbiter mission.\\
   {Our aims are to demonstrate the feasibility of regularized imaging spectroscopy as a method for analysis of STIX data, and also to show how such analysis can lead to insights into the physical processes affecting the nonthermal electrons responsible for the hard X-ray emission observed by STIX. \\
   {STIX records imaging data in an intrinsically different manner from RHESSI.  Rather than sweeping the angular frequency plane in a set of concentric circles (one circle per detector), STIX uses $30$ collimators, each corresponding to a specific angular frequency. In this paper we derive an appropriate modification of the previous computational approach for the analysis of the visibilities observed by STIX. This approach also allows for the observed count data to be placed into non-uniformly-spaced energy bins.\\
   {We show that the regularized imaging spectroscopy approach is not only feasible for analysis of the visibilities observed by STIX, but also more reliable. Application of the regularized imaging spectroscopy technique to several well-observed flares reveals details of the variation of the electron flux spectrum throughout the flare sources.\\
   {We conclude that the visibility-based regularized imaging spectroscopy approach is well-suited to analysis of STIX data. We also use STIX electron flux spectral images to track, for the first time, the behavior of the accelerated electrons during their path from the acceleration site in the solar corona toward the chromosphere.\\
   
\textbf{key words.} Techniques: imaging spectroscopy; Sun: X-rays, gamma rays; Sun: flares; Methods: numerical



\section{Introduction}\label{S-Introduction}

Imaging spectroscopy in solar hard X-rays, i.e., the construction of spatio-spectral cubes describing a flaring source in the energy range between a few and a few hundred~keV, was the central objective of the NASA Reuven Ramaty High-Energy Solar Spectroscopic Imager (RHESSI) mission \citep{2002SoPh..210....3L}. Because of its rotating modulation collimator design \citep{1965ApOpt...4..143O}, RHESSI native measurements were in the form of two-dimensional spatial Fourier components of the incoming photon flux, termed \emph{visibilities} \citep{2002SoPh..210...61H}, and this inherent aspect of the data drove the formulation, implementation, and application of several innovative approaches to imaging spectroscopy for RHESSI. Notable among these approaches was one \citep{2007ApJ...665..846P,prato2009regularized} that combined regularized spectral inversion and Fourier-based image reconstruction methods to produce images of the flux of bremsstrahlung-producing electrons at different energies or, equivalently, electron flux spectra at different points in the image.

The \cite{2007ApJ...665..846P} approach exploits the linearity of both spatial Fourier transform and photon $\rightarrow$ electron spectral inversion operations to reverse the conventional order of spatial and spectral inversions. Instead of performing a spatial Fourier transform on the observed count visibilities to produce count images, using these images to produce count spectra at different locations, and then spectrally inverting the count spectra at each location, the method reverses the order of spatial and spectral processing. It takes the observed count visibilities and, in conjunction with an assumed bremsstrahlung cross-section \citep[e.g.,][]{1959RvMP...31..920K}, performs a regularized spectral inversion procedure (all in the spatial-frequency domain) to arrive at the visibilities associated with the flux spectrum of the electrons responsible for the bremsstrahlung emission. Due to the regularization \citep{pianabook} inherent in their construction, these electron visibilities vary smoothly from one electron energy channel to the next, and it must be noted that this property is \emph{not} shared by the native count visibilities. Count visibilities are constructed from independent bundles of detected counts in each prescribed energy range and, since each bundle of counts has its own independent level of statistical noise, the measured count number can vary erratically from one count energy channel to the next. This feature is exacerbated by the (ill-posed) nature of the spectral inversion problem \citep{1986ipag.book.....C}, resulting in inferred electron spectra that can possess large, erratic (and doubtless unphysical) variations with energy, and are hence of little use in addressing physical issues related to electron acceleration and transport.

Spatial inversion, via a discrete Fourier transform, of the electron visibilities yields images of the electron flux at different electron energies $E$, and, because of the smooth energy variation of the electron visibilities from which they are constructed, these images also vary smoothly with electron energy. This allows a more reliable (and indeed more sophisticated) physical investigation into the variation of the electron spectrum from point to point within the flare. Among the scientific results provided by this approach using RHESSI data, we mention the determination of the acceleration region size in flaring loops \citep{2012A&A...543A..53G}, of the number of particles within the flare acceleration region \citep{2012ApJ...755...32G}, of the values of the emission filling factor and the specific acceleration rate \citep{2013ApJ...766...28G}, and of the parameters associated with stochastic acceleration models \citep{2010ApJ...712L.131P}. 

In the present era, visibility-based observations of solar flare hard X-ray emission are obtained with the Spectrometer/Telescope for Imaging X-rays \citep{krucker2020spectrometer}, one of the six remote-sensing instruments on board the ESA cluster on the Solar Orbiter mission \citep{2020A&A...642A...1M}. In contrast with the rotating modulation collimator approach of RHESSI, which generated visibilities by means of a rather involved data-stacking process \citep{2002SoPh..210...61H}, STIX computes visibilities in a completely static way, at a fixed set of points in the spatial- (or more correctly, angular-) frequency domain, using the properties of the Moir\'e patterns produced by pairs of grids with slightly different geometry \citep{2023SoPh..298..114M}. Although, for a given flare, the number of measured RHESSI visibilities could in principle have been quite large, in practice the signal in many cases was not sufficient to guarantee a reliable visibility measurement, particularly at higher energies, with their lower count rates. Furthermore, the spatial resolution of the instrument was compromised by blurring effects introduced by the data stacking process. By contrast, each one of the $30$ STIX visibilities is provided by a unique STIX collimator, so that all visibilities are observed at all time points and at all energies, without any blurring effects. 

The first objective of the present work is to provide an implementation of the regularized imaging spectroscopy method that was introduced for RHESSI, but which is now optimized for the STIX framework discussed above. We have also chosen, in contrast to the algorithm developed for RHESSI, to design an algorithm that can be applied to non-uniformly-distributed energy channels. By considering a specific case study, we illustrate the STIX high-level research products that can be generated by means of regularized imaging spectroscopy; we discuss how the reliability of such products can be validated; and we provide a quantitative estimate of the higher accuracy of STIX electron flux spectral images compared to the ones obtained from RHESSI. Finally, we use this regularized imaging spectroscopy approach to show that, with STIX observations, the spectral behavior of accelerated electrons in solar flares can be tracked from the corona to the chromosphere, with associated implications for the nature of electron transport mechanisms in solar flares. 

The plan of the paper is as follows. Sect.~\ref{S-methodology} summarizes the regularized imaging spectroscopy approach in the case of visibility-based telescopes, while Sect.~\ref{S-application} applies this approach to experimental measurements recorded by STIX. Sect.~\ref{S-5-8-21} discusses preliminary results obtained with the regularized imaging spectroscopy technique in the case of a solar flare recorded on May 8, 2021 (SOL2021-05-08T18).
Our conclusions are offered in Sect.~\ref{S-conclusions}.

\section{Electron flux spectral images}\label{S-methodology}

We here follow Appendix~A of \cite{2007ApJ...665..846P} in reviewing the various concepts, and related quantities, that are involved in the construction of images (maps) of the electron flux at different energies from STIX observations of count visibilities.

\subsection{Underlying concepts}\label{sec:underlying-math}

Define a Cartesian coordinate system $(x,y,z)$ such that $(x,y)$ (each in arcsecond units) represents a location in the image plane and $z$ represents distance (measured in cm) along the line of sight. Let the local density of target particles within a source of line-of-sight depth $\ell(x,y)$ (cm) be $n_{\rm target}(x,y,z)$ (cm$^{-3}$) and let the  electron flux spectrum, differential in energy $E$, at the point $(x,y,z)$ in the source be $F(x,y,z; E)$ (electrons~cm$^{-2}$~s$^{-1}$~keV$^{-1}$). We formally define the \emph{mean electron flux spectrum} ${\overline F}(x,y; E)$ (electrons~cm$^{-2}$~s$^{-1}$~keV$^{-1}$) by \citep{pianabook}

\begin{eqnarray}\label{eq:fbar-definition}
{\overline F} (x,y;E) &=& \frac{\int_{z=0}^{\ell(x,y)} n_{\rm target}(x,y,z) \, F(x,y,z;E) \, dz} {\int_{z=0}^{\ell(x,y)} n_{\rm target}(x,y,z) \, dz} \cr
&\equiv& \frac{\int_{z=0}^{\ell(x,y)} n_{\rm target}(x,y,z) \, F(x,y,z;E) \, dz}{{\overline n}_{\rm target}(x,y) \, \ell(x,y)} \, \,\,\, ,
\end{eqnarray}
where ${\overline n}_{\rm target}(x,y)$ is the density of target protons, averaged along the line of sight distance $\ell(x,y)$. The corresponding \emph{photon spectrum image} $I(x,y;\epsilon)$ (photons~cm$^{-2}$~s$^{-1}$~keV$^{-1}$~arcsec$^{-2}$) produced by bremsstrahlung encounters of these electrons with ambient protons is \citep{2003ApJ...595L.115B}

\begin{eqnarray}\label{eq:fundamental}
&I(x,y; \epsilon) & \hspace{-0.3cm} = \frac{a^2}{4 \pi R^2} \cdot \cr
& \hspace{0.9cm}& \cdot  \int^{\infty}_{E =\epsilon}  \int_{z=0}^{\ell(x,y)} n_{\rm target}(x,y,z) \, F(x,y,z; E) \, Q(\epsilon, E) \, dE \, dz  \cr
&=& \hspace{-0.6cm} \frac{a^2 }{4 \pi R^2}  {\overline n}_{\rm target}(x,y) \, \ell (x,y) \int^{\infty}_{E =\epsilon} {\overline F}(x,y; E) \, Q(\epsilon, E) \, dE \, \,\,\,,
\end{eqnarray}

where $R$ (cm) is the distance from the source to the instrument, $Q(\epsilon, E)$ (cm$^2$~keV$^{-1}$) is the bremsstrahlung cross-section, differential in photon energy $\epsilon$, for photon emission at energy $\epsilon$, and $a = 7.25 \times 10^7 \, {\widetilde R}$~cm~arcsec$^{-1}$ is the conversion factor from arcseconds to cm, ${\widetilde R}$ being the distance from the source to the observer in astronomical units (au). Eq. \eqref{eq:fundamental} may be written

\begin{equation}\label{eq:fundamental-1}
I(x,y; \epsilon) = \frac{1}{4 \pi R^2} \, \int^{\infty}_{E =\epsilon} \mathcal{F}(x,y;E) \, Q(\epsilon, E) \, dE \,\,\, ,
\end{equation}
where we have defined the \emph{electron flux spectral image} (electrons~cm$^{-2}$~s$^{-1}$~keV$^{-1}$~arcsec$^{-2}$)

\begin{equation}\label{eq:electron-map}
\mathcal{F}(x,y;E) = a^2 \,\, {\overline n}_{\rm target}(x,y) \, \ell(x,y) \, {\overline F}(x,y; E) \,\,\, .
\end{equation}
We can treat the set of electron flux spectral images either as maps of the electron flux at a prescribed electron energy $E$, or as a set of mean electron flux spectra at each point $(x,y)$ in the image plane. 

Given that the STIX instrument on Solar Orbiter observes the Sun from different distances at different times in its heliocentric orbit, a brief digression into the role of the source-observer distance $R$ is pertinent, and also yields some valuable insights. For a given target density $n_{\rm target}(x,y,z)$ and electron flux $F(x,y,z;E)$ profile, the observed photon spectrum image, i.e., the hard X-ray emission from a region of prescribed angular extent, is, according to Eq.~\eqref{eq:fundamental}, \emph{independent of the source-observer distance $R$}. At a smaller (say) value of $R$, a given solid  angle extent (in arcseconds$^2$) corresponds to a smaller area on the source plane. The cm$^2$-to-arcsecond$^2$ conversion factor $a^2$ is accordingly smaller, leading to the multiplicative factor $a^2/4 \pi R^2$ being independent\footnote{It could also be noted that the quantity $a^2/4 \pi R^2$ is equal to $(1/206265)^2/4 \pi$, i.e., the solid angle corresponding to one square arcsecond divided by the entire $4\pi$ steradian sphere, and so is straightforwardly independent of the source-observer distance.} of $R$. Physically, a smaller number of photons are emitted from the reduced physical area corresponding to a square arcsecond of angular extent. However, this factor is exactly cancelled by the reduced source-observer distance in the inverse-square law relating the number of emitted photons and the number that impact a distant detector. Thus, while a given flare event will result in more recorded counts for an instrument located at ${\widetilde R} < 1$ than for an identical instrument at $1$~au, this is because the angular extent of the source is larger, and not because the photon spectrum image $I(x,y; \epsilon)$ is any brighter. 

\subsection{Determining electron flux spectral images from STIX observations}\label{sec:electron-flux-maps}

The STIX instrument \citep{krucker2020spectrometer} on the Solar Orbiter mission \citep{2020A&A...642A...1M} does not directly image hard X-rays. 
Rather, a set of $30$~segmented detectors, each located behind a pair of absorbing grids with different geometrical properties \citep{2023SoPh..298..114M}, record count measurements, in each of four detector segments, at different energies. 
This information is then combined to obtain the values of $30$~two-dimensional spatial Fourier components, or \emph{visibilities}, of the source structure at the count energy in question.
Each visibility corresponds to a specific angular frequency $(u,\varv)$, where $u$ and $\varv$ are in units of arcsec$^{-1}$ \citep[see Fig.~10 of][]{2023SoPh..298..114M}.


The relationship between a \emph{count visibility} $V(u,\varv;q)$ (counts~cm$^{-2}$~s$^{-1}$~keV$^{-1}$), recorded at energy $q$, and the photon images $I(x,y;\epsilon)$ (photons~cm$^{-2}$~s$^{-1}$~keV$^{-1}$~arcsec$^{-2}$) at different photon energy values $\epsilon$ is given by \citep{2007ApJ...665..846P}

\begin{equation}\label{eq:vdef_app_j}
V(u,\varv;q) = \int_X \, \int_Y \, \int_{\epsilon = q}^\infty \, D(q, \epsilon) \, I(x,y;\epsilon) \, e^{2 \pi i(ux + \varv y)} \, d \epsilon \, dx \, dy ~,
\end{equation}
where the spatial integrals extend over the entire field of view of the instrument, and $D(q,\epsilon)$ (keV$^{-1}$) is the value of the detector response matrix\footnote{Note the distinction between Eq.~\eqref{eq:vdef_app_j} and Eq.~(A5) of \cite{2007ApJ...665..846P}, in which $D(q,\epsilon)$ is taken to be dimensionless and a corresponding elementary energy range $dq$ appears on the left side.} corresponding to the probability per unit count energy of a count with energy $q$ being produced by a photon of energy $\epsilon$. We note that while the distance from the source to an instrument  does not change the magnitude of the photon image for a given event (since, as discussed in Sect.~\ref{sec:underlying-math}, the photon image is defined as the emissivity per arcsec$^2$ on the sky, and at a smaller source-observer distance a specified angular size corresponds to a smaller physical distance on the source), the field of view of STIX is sufficiently large that observed visibilities measure the emission from the entire Sun, and hence the magnitude of the visibilities $V(u, \varv; q)$ \emph{do} increase as the distance from the Sun to the instrument decreases.

Combining Eqs.~\eqref{eq:fundamental-1} and~\eqref{eq:vdef_app_j} leads to the mathematical expression that links the electron flux spectral images $\mathcal{F}(x,y;E)$ to the STIX count visibilities $V(u, \varv;q)$, namely \citep[cf. Eq.~(A6) of][]{2007ApJ...665..846P}

\begin{equation}\label{eq:count-visibilities-electron-maps}
\begin{split}
&V(u, \varv;q)  = \frac{1}{4 \pi R^2} \cdot \\
&\cdot  \int_X \, \int_Y \, \int_{\epsilon=q}^\infty \, \int_{E =\epsilon}^{\infty} \mathcal{F}(x,y;E) \, D(q,\epsilon) \, Q(\epsilon, E) \, e^{2 \pi i(ux + \varv y)} \, dE \, d \epsilon \, dx \, dy  \,\,\, .
\end{split}
\end{equation}
Reversing the order of integration over~$\epsilon$ and~$E$ gives

\begin{eqnarray}\label{eq:main-result}
&V(u, \varv;q)  & \hspace{-0.2cm}= \frac{1}{4 \pi R^2} \cdot \cr
& & \hspace{-1.5cm}\cdot  \int_X \, \int_Y \, \int_{E =q}^{\infty} \int_{\epsilon=q}^E \, \mathcal{F}(x,y;E) \, D(q,\epsilon) \, Q(\epsilon, E) \, e^{2 \pi i(ux + \varv y)} \, d \epsilon \, dE \, dx \, dy \cr
&\hspace{-0.55cm}=& \hspace{-0.8cm} \frac{1}{4 \pi R^2} \int_X \, \int_Y \int_{E = q}^{\infty} \mathcal{F}(x,y;E) \, K(q,E) \, e^{2 \pi i(ux + \varv y)} \, dE \, dx \, dy \,\,\, ,
\end{eqnarray}
where we have defined the \emph{count cross section}



\begin{equation}\label{eq:ce_cross}
K(q,E) = \int_{\epsilon = q}^E D(q,\epsilon) \, Q(\epsilon, E) \, d \epsilon \,\,\, ,
\end{equation}
with units of cm$^2$~keV$^{-1}$, differential in count energy $q$. This quantity combines the bremsstrahlung cross-section for production of a photon of energy $\epsilon$ by an electron of energy $E$ with the probability (per unit count energy) that such a photon will be recorded as a count with an energy $q$, to give the overall cross-section, differential in count energy, for production of a count of energy $q$ by an electron of energy $E$.

Finally, for each $(u,\varv)$ we introduce the \emph{electron flux visibility spectrum} (electrons~cm$^{-2}$~s$^{-1}$~keV$^{-1}$)

\begin{equation}\label{eq:evis_app}
W(u,\varv;E) = \int_X \, \int_Y \, \mathcal{F}(x,y; E) \,
e^{2 \pi i(ux + \varv y)} \, dx \, dy \,\,\, ,
\end{equation}
the Fourier transforms of the electron flux spectral image at the measured spatial frequencies $\{(u_i,\varv_i)\}_{i=1}^{30}$ and electron energy $E$. Then, changing the order of integration over space and electron energy in the last equality of Eq.~\eqref{eq:main-result}, we finally derive the relationship between the observed differential count visibility spectrum $V(u,\varv;q)$ and the underlying electron flux visibility spectrum:

\begin{equation}\label{eq:result_red}
V(u,\varv;q) = \frac{1}{4 \pi R^2} \, \int_q^{\infty} W(u,\varv;E) \, K(q,E) \, dE \,\,\, .
\end{equation}
Eq.~\eqref{eq:result_red} is the fundamental result presented in \cite{2007ApJ...665..846P}, and its similarity to Eq.~\eqref{eq:fundamental} essentially shows that the spectral relationship between count and electron quantities in the spatial domain carries through to the spatial frequency domain. Electron flux visibility spectra $W(u, \varv;E)$, which contain information on the spatial (more accurately, angular on the plane of the sky) distribution of the electron flux spectral images are not, of course, directly measured by STIX, but they can be retrieved from the observed count visibility spectra $V(u, \varv; q)$ by inverting Eq.~\eqref{eq:result_red} via any number of established inversion techniques \citep{pianabook}. Specifically, for a given angular frequency $(u,\varv)$, the (complex) values of $W(u,\varv; E)$ at different energies can be retrieved by applying a well-tested Tikhonov regularization approach \citep{1994A&A...288..949P,2003ApJ...595L.127P} to both the real and the imaginary parts of Eq.~\eqref{eq:result_red}. We refer the reader to \cite{2007ApJ...665..846P} and to \cite{prato2009regularized} for more details about the implemented inversion method.

The electron flux spectral images for different energies $\mathcal{F}(x,y;E)$ can be constructed from the set of electron flux visibility spectra $\{W(u_i,\varv_i; E)\}_{i=1}^{30}$ by applying standard image reconstruction methods \citep[e.g.,][]{pianabook}. We remind the reader that as a result of the regularized spectral inversion procedure used to generate the electron flux visibilities $W(u, \varv; E)$ these visibilities, and hence the electron flux spectral images $\mathcal{F} (x, y; E)$ that are constructed from them, vary smoothly with energy $E$, unlike the count visibilities $V(u, \varv; q)$ on which they are based. As we shall see, this feature of the electron flux spectral images permits a more impactful scientific analysis than is possible from the study \citep[e.g.,][]{2008ApJ...673..576X} of count-based images.

\begin{figure*}[t]
\centering
\includegraphics[width=0.9\textwidth]
{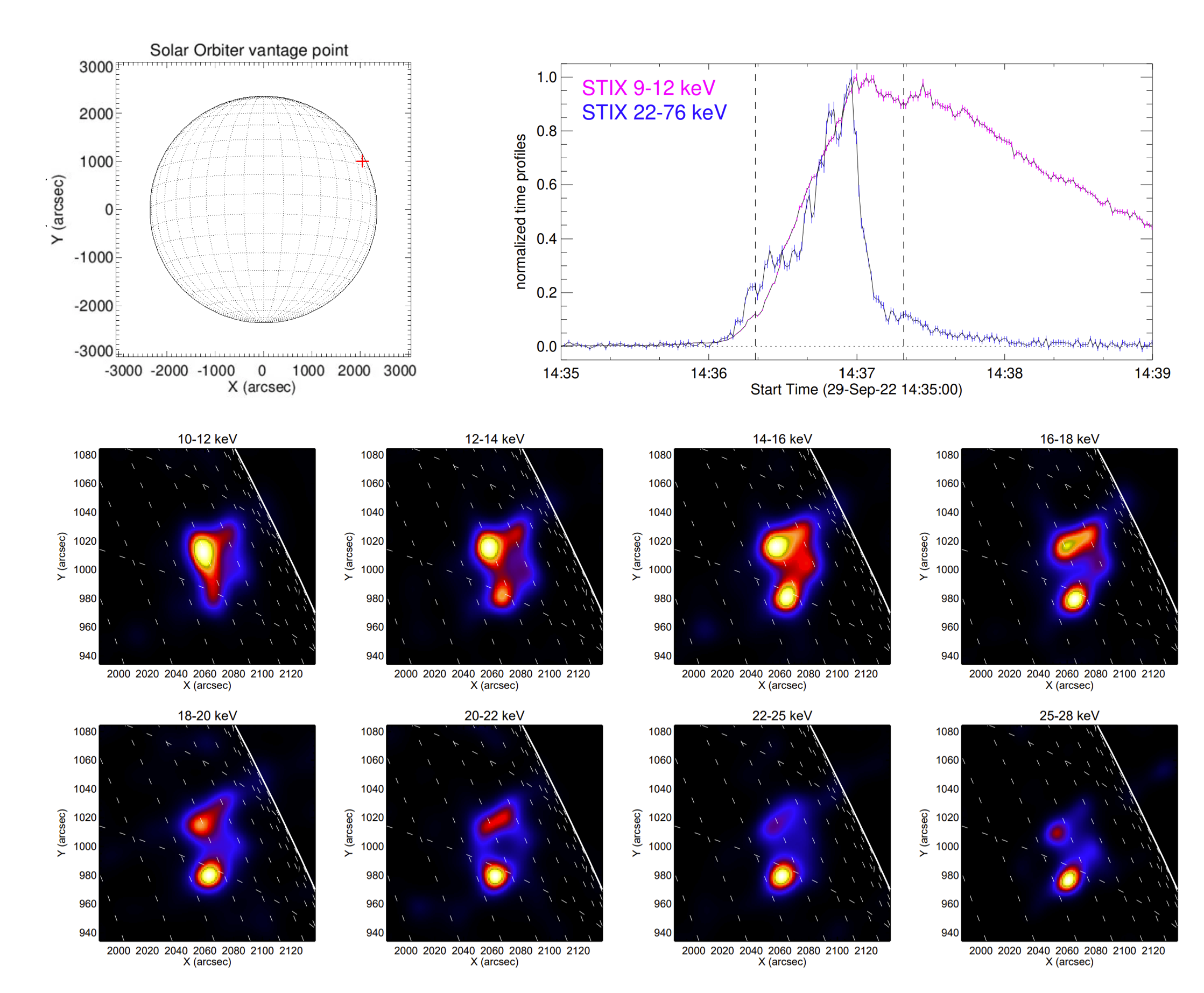}
\caption{
The SOL2022-09-29T14 event.
\emph{Top row, left panel}: position of the event on the solar disk; \emph{right panel}: normalized STIX time profiles corresponding to the energy ranges 9-12~keV and 22-76~keV. The reported time is relative to Solar Orbiter.
\emph{Bottom row}: count images of the event at different energies as reconstructed by MEM$\_$GE.}
\label{fig:event}
\end{figure*}

As noted earlier, this visibility-based imaging spectroscopy methodology was originally developed in the RHESSI framework. Inspired by that approach, here we developed an algorithm that is tailored to the use of STIX visibilities. From a technical viewpoint, there are two main differences between the two methods. First, for RHESSI, at some $(u, \varv)$ points the count visibilities $V(u, \varv;q)$ often suffered gaps at specific $q$ values, as a consequence of the sub-optimal data stacking process that characterizes the underlying rotating modulation collimator (RMC) imaging concept; instead, for STIX there is, by design, a one-to-one correspondence between sampled visibilities and collimators, so that count visibility spectra are always complete. {Second, the implementation of the algorithm realized for RHESSI constrained the count visibility spectra to be uniformly sampled along the count energy direction. Instead, the STIX algorithm considers non-uniformly-spaced count energy channels by introducing appropriate weights in the quadrature formula for the discretization of Eq.~\eqref{eq:result_red} and by accounting for such weights in the regularized inversion formula.

\section{Application to STIX visibilities}\label{S-application}

In order to assess the electron flux spectral image methodology using STIX visibilities, we considered a case study represented by the near-limb (as viewed from Solar Orbiter) flare observed by STIX on September 29, 2022 (SOL2022-09-29T14).
Although this flare occurred on the far side of the Sun as seen from Earth, from STIX observations the estimated GOES class is $\sim$M1, with an uncertainty range from C7 up to~M2. The top row panels of Fig.~\ref{fig:event} show the position of the source on the solar disk and the (normalized) STIX time profiles corresponding to the energy ranges 9-12 keV and 22-76 keV. The two bottom rows contain the count images reconstructed by applying MEM$\_$GE, a constrained maximum entropy algorithm \citep{Massa_2020} to the visibility bag corresponding to the time interval between 14:36:19 and 14:37:19~UT (relative to Solar Orbiter). As we now discuss, the regularized imaging spectroscopy approach allows users to obtain research products of higher level than simply count images; they include regularized electron flux spectral images, regularized count images, and spatially-resolved regularized count and electron flux spectra.

\subsection{High-level research products}

\begin{figure*}[t]
\centering
\includegraphics[width=\textwidth]{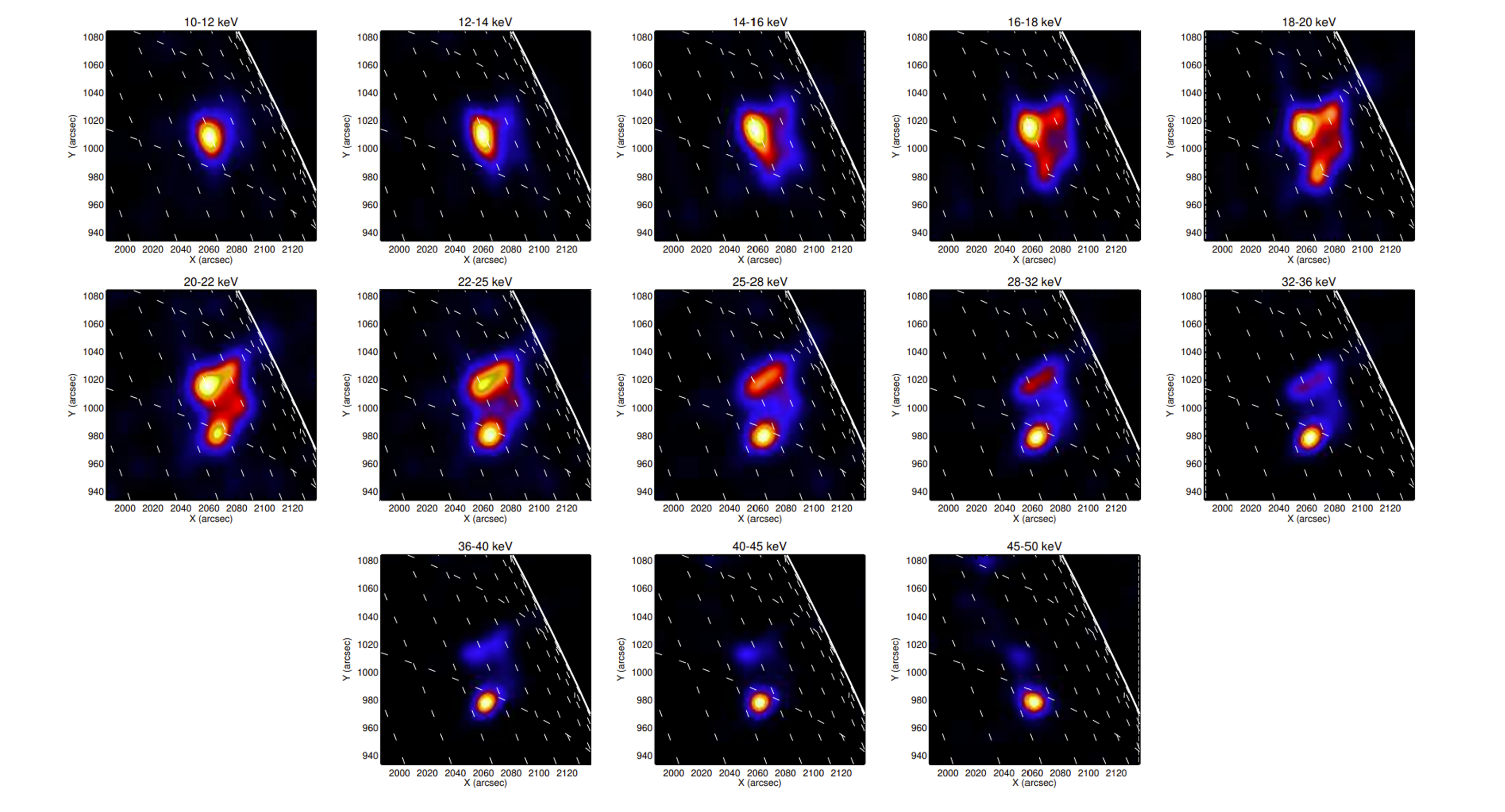}
\caption{Electron flux images in different energy channels for the 
SOL2022-09-29T14 event, as reconstructed by MEM$\_$GE.}
\label{fig:electron-maps}
\end{figure*}

The algorithm described in the previous section, using the isotropic form of the bremsstrahlung cross-section in \cite{1959RvMP...31..920K}, provides regularized electron flux visibility spectra for each $(u, \varv)$ point sampled by STIX. These data can be re-ordered by collecting together all values of the regularized solutions $W(u, \varv;E)$ of Eq.~\eqref{eq:result_red} corresponding to the same electron energy $E$. The application of the MEM$\_$GE algorithm\footnote{Similar reconstructions can be obtained by applying other visibility-based reconstruction algorithms; see, e.g.,  \cite{2022A&A...668A.145V,2023ApJS..268...68P,2023SoPh..298..114M}.} to each of these electron flux visibilities provides images of the electron flux at different energies, as shown in Fig.~\ref{fig:electron-maps}. As a consequence of the regularization process involved in constructing the electron flux visibilities used to produce the electron flux spectral images, these images necessarily vary smoothly with electron energy. Furthermore, the algorithm provides visibilities at electron energies that are larger than the highest count energy observed in the count images of Fig.~\ref{fig:event}; this is due to the fact that counts registered in a specific energy channel result from photons that are in turn produced via bremsstrahlung by electrons with all energies higher than that of the photon, so that counts at a given energy provide information on electrons of all higher energies. 

A second high-level research product can be obtained by substituting the (regularized) electron visibility spectra into the bremsstrahlung-like equation~(\ref{eq:result_red}) to produce count visibility spectra \emph{that are also regularized}. Applying an inverse spatial Fourier transform to these produces the \emph{regularized count images} of Fig.~\ref{fig:parametric-maps}. Unlike the raw count-based images, these regularized images have the desirable property of varying smoothly with (count) energy. Further, the count cross section $K(q,E)$ in Eq.~\eqref{eq:result_red} can be replaced with the bremsstrahlung cross section $Q(\epsilon,E)$ to obtain \emph{regularized photon visibilities} and hence, via application of a spatial inverse Fourier transform, \emph{regularized photon images}. For these products, the instrumental dependency is removed.

\begin{figure*}[t]
\centering
\includegraphics[width=0.85\textwidth]
{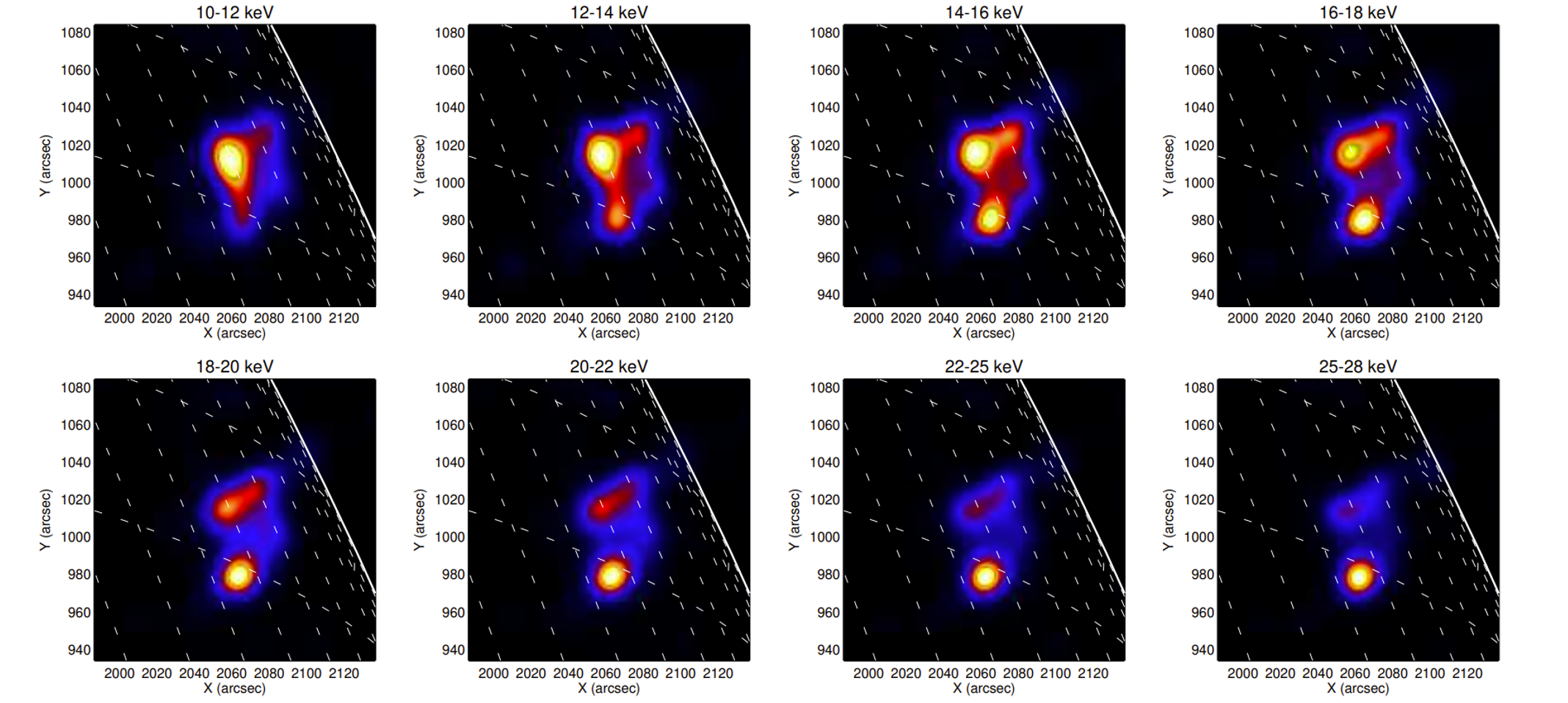}
\includegraphics[width=1.\textwidth] {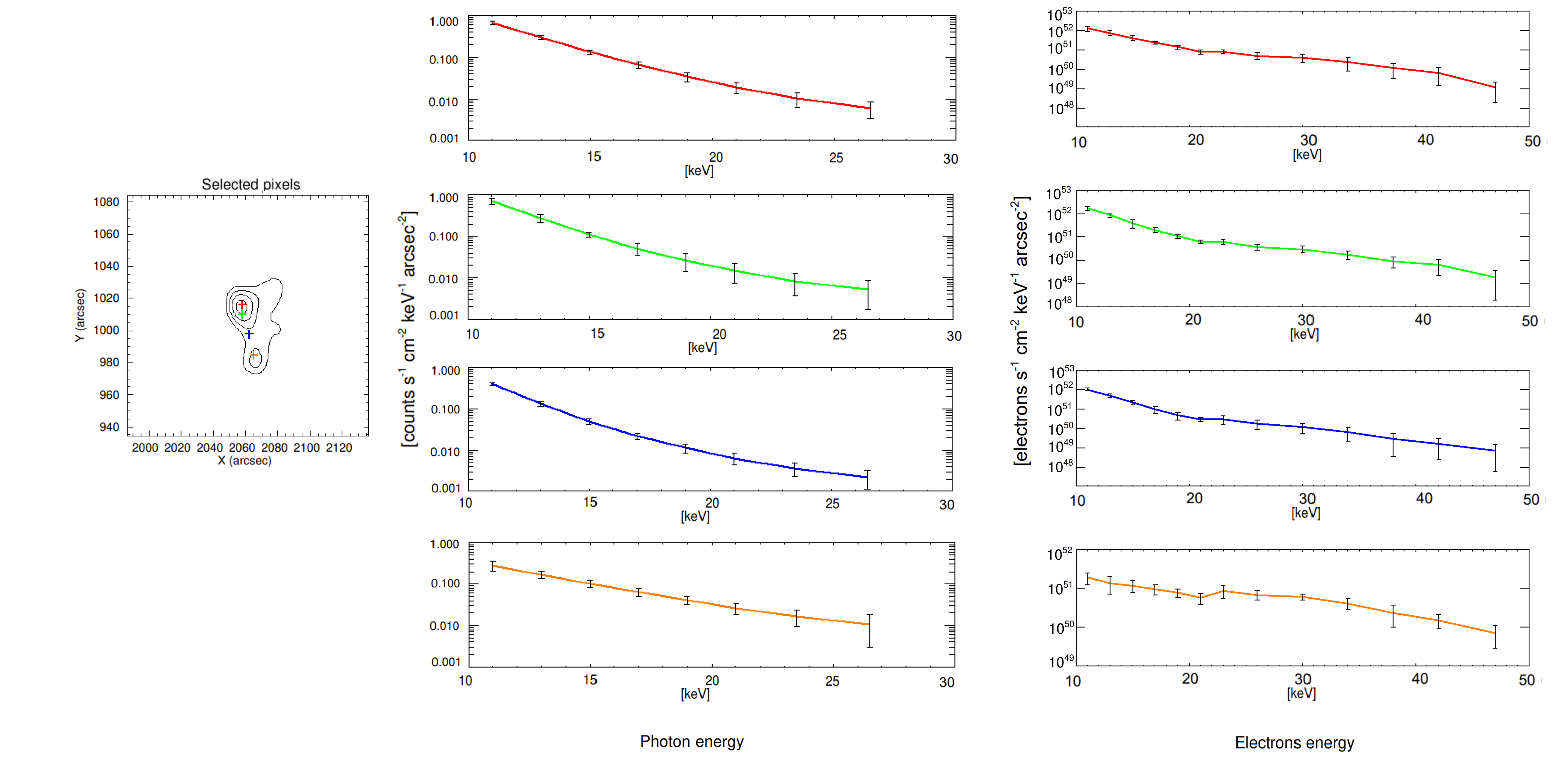}
\caption{\emph{Top panels}: Regularized count images in different energy channels for the SOL2022-09-29T14 event, as reconstructed by MEM$\_$GE. \emph{Bottom left panel}: selected locations are indicated with colored crosses. \emph{Middle column}: spectra at different locations obtained from the regularized count images. \emph{Right column}: pixel-wise spectra obtained from the regularized electron flux spectral images.}
\label{fig:parametric-maps}
\end{figure*}

Perhaps the most sophisticated high-level research products provided by regularized imaging spectroscopy are the ones contained in Figs.~\ref{fig:parametric-maps} and~\ref{fig:local-spectra}. Taking into account the limited spatial resolution of STIX, we average the values of the regularized count images and electron flux spectral images over areal boxes of $5$~arcsec $\times \, 5$~arcsec size and then construct regularized count and electron spectra for each such $25$~arcsec$^2$ area. The spectrum at each such location can then be analyzed, e.g., by fitting with parametric models. Thanks to the smoothing effects of regularization, this fitting process is significantly more reliable than the one applied to non-regularized spectra; indeed, the regularized spectra suffer less spurious oscillations and are characterized by uncertainties that are significantly smaller than the ones associated with non-regularized spectra.

Below are a few comments on the results presented in the third column of the bottom panels of Fig.~\ref{fig:parametric-maps}. 

\begin{itemize}

    \item For the northern footpoint (indicated by a red plus sign in the left panel of Fig.~\ref{fig:parametric-maps}), the corresponding panel of Fig.~\ref{fig:parametric-maps} clearly indicates both thermal (below 20~keV) and non-thermal (above 20~keV) components of the electron flux spectrum, with the transition between these regimes evidenced by the spectral flattening above 20~keV.
    
    \item The area highlighted by a green plus sign corresponds to an area shifted by 10~arcsec towards the South relative to the northern source. Its spectrum shows a very similar behavior; however, as the image angular resolution is \(\sim\)14 arcsec, the spectra for these two regions are not completely independent of each other.
    
    \item The spectrum associated with the southern source (indicated with an orange plus sign in the left panel of Fig.~\ref{fig:parametric-maps}) is harder to interpret. As it can be seen in Fig.~\ref{fig:electron-maps}, this source is visible in the electron flux images only above 16~keV and, therefore, the spectral points below this energy are not reliably determined.

    \item The spectrum corresponding to the area in between the northern and southern sources is indicated with a blue plus sign. At this location, the electron flux spectrum is similar to the other sources indicated; however, considerations of spatial resolution and dynamic range demand that caution must be taken in the interpretation of this spectrum. While the STIX dynamic range can be as large as $\sim$20:1 for a single strong source \citep[see Sect.~4.3 of][]{krucker2020spectrometer}, limited data statistics and complex source morphology reduce this to a value that is typically $\sim$10:1 for images such as those presented in Figs.~\ref{fig:event} through~\ref{fig:parametric-maps}. Only clearly identifiable sources within the dynamic range of the STIX instrument in the considered energy range can be reliably be used for scientific analyses, and the electron flux images of Fig.~\ref{fig:electron-maps} do not show a clearly identifiable source at this location. Thus, the inferred electron spectrum in this region is likely a result of the influence of the other nearby sources in in the image.

\end{itemize} 

Similar considerations apply to the regularized count spectra shown in the second column of the bottom panels of Fig.~\ref{fig:parametric-maps}.

\subsection{Validation}\label{SS-validation}

\begin{figure*}
    \centering
    \includegraphics[width=0.9\textwidth]{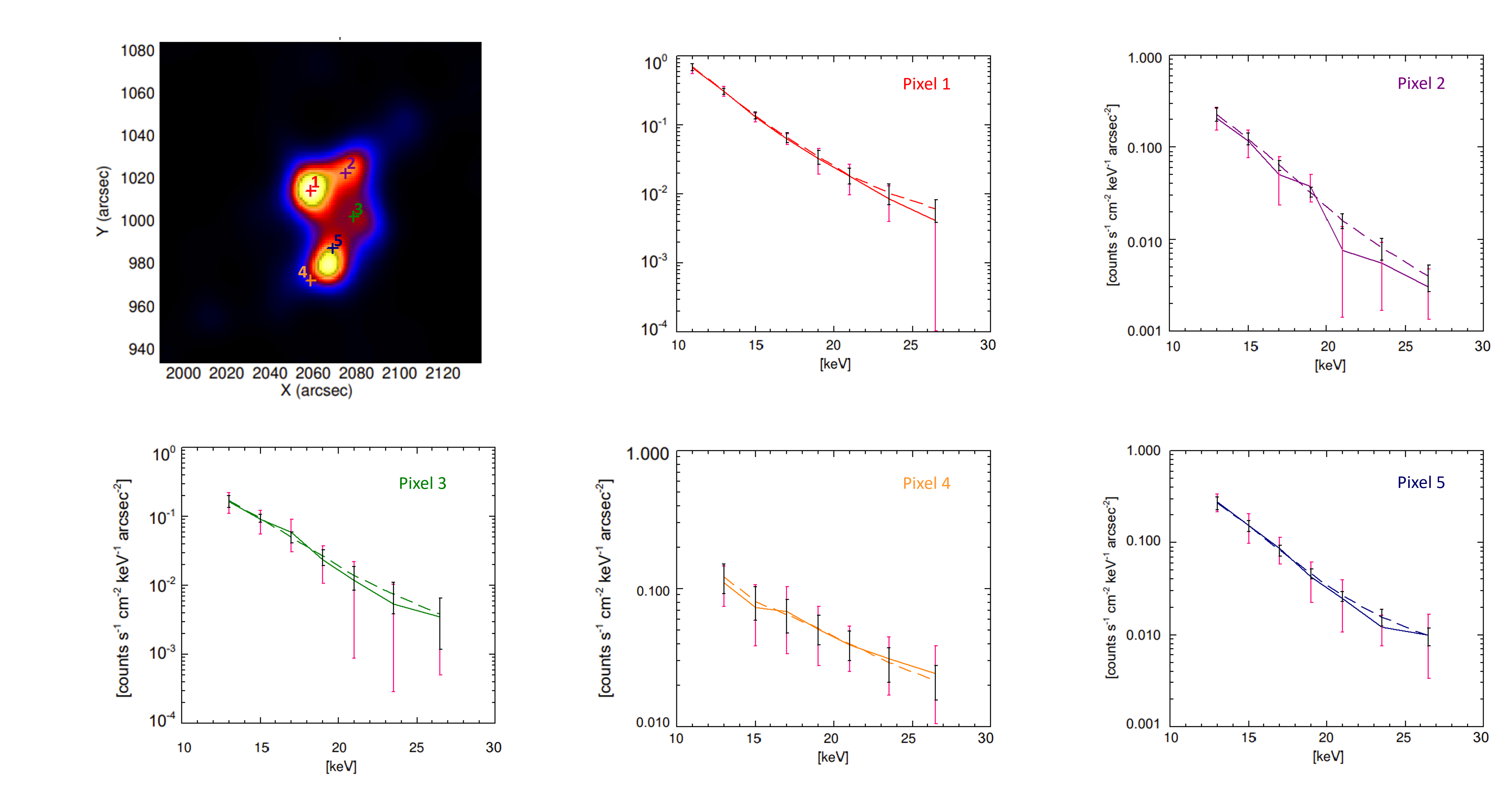}
    \caption{Spatially resolved count spectra for the SOL2022-09-29T14 event.
    \emph{Top left panel}: Selected pixels are indicated with colored crosses. Other panels show the corresponding pixel-wise spectra obtained from count images (solid line) and regularized count images (dashed line) as a function of count energy. The pixels selected in the top left panel and their respective spectra are indicated with the same color. The plots are logarithmically scaled on the $y$-axis. The ($3\sigma$) error bars on the count spectra are in pink, while the error bars on the regularized count spectra are in black.}
    \label{fig:local-spectra}
\end{figure*}

\begin{table*}
\small
\centering
\begin{center}
\begin{tabular}{ccccc}
\toprule
 Event & OSPEX & electron flux spectral images & count images & regularized count images \\
\midrule
 SOL2021-05-08T18
 & $\gamma = 5.28 \pm 0.18$ & $\delta = 4.53 \pm 0.07$ &  $\gamma = 4.94 \pm 0.24$ &  $\gamma = 5.50 \pm 0.04$  \\
 \midrule
  SOL2021-08-26T23
& $\gamma = 5.65 \pm 0.12$ & $\delta = 4.59 \pm 0.11$ &  $\gamma = 5.27 \pm 0.17$ &  $\gamma = 5.57 \pm 0.06$  \\
 \midrule
 SOL2022-01-20T05
 & $\gamma = 6.51 \pm 0.26$ & $\delta = 5.07 \pm 0.03$ &  $\gamma = 6.25 \pm 0.14$ &  $\gamma = 6.36 \pm 0.01$  \\
  \midrule
 SOL2022-08-28T16
 & $\gamma = 6.74 \pm 0.18$ & $\delta = 4.97 \pm 0.04$ &  $\gamma = 6.88 \pm 0.24$ &  $\gamma = 6.81 \pm 0.03$  \\
   \midrule
 SOL2022-09-29T14
 & $\gamma = 4.54 \pm 0.05$ & $\delta = 3.68 \pm 0.01$ &  $\gamma = 4.24 \pm 0.11$ &  $\gamma = 4.42 \pm 0.01$  \\
 \bottomrule
 \end{tabular}
 \end{center}
 \caption{\emph{First column:} event date. \emph{Second column:} photon power-law index $\gamma$ obtained by fitting spatially integrated spectra with OSPEX. \emph{Third column:} electron power-law index $\delta$ obtained by fitting spatially integrated electron flux spectral images. \emph{Fourth and fifth columns:} photon power-law index $\gamma$ obtained by fitting spatially integrated count images and spatially integrated regularized count images, respectively, assuming a unit detector response matrix.  The uncertainties in columns 3,~4, and~5 are determined by repeating the computation for 10~random realizations of the input visibilities.}
 \label{tab:spectral-indices}
\end{table*}

In order to assess the validity and reliability of the regularized imaging spectroscopy method, we studied the impact that regularization has on both spatially-resolved and spatially-integrated spectra. Using the same 
SOL2022-09-29T14 event as the case study, Fig.~\ref{fig:local-spectra} considers the five positions in the flare highlighted in the top left panel and, for each one, compares the local count spectra provided by the original images in the bottom rows of Fig.~\ref{fig:event} with the ones provided by the regularized count images in Fig.~\ref{fig:parametric-maps}. The five plots clearly show that while regularization does not modify the overall shape of the local count spectra, it does increase their stability, in particular for locations in the flaring source where the signal-to-noise ratio is smaller.

For thin-target hard X-ray emission (appropriate for calculating the count spectrum image $J(x,y;q)$ (counts cm$^{-2}$~s$^{-1}$~keV$^{-1}$~arcsec$^{-2}$) produced by a mean electron flux spectrum $F(x,y;E)$ at a given position $(x,y)$), use of relatively simple (e.g., Kramers, Bethe-Heitler) forms of the bremsstrahlung cross-section, and assuming a diagonal Detector Response Matrix $D(q,\epsilon)$ with constant entries, lead to a simple approximate relation between the power-law spectral indices $\gamma$ and $\delta$ of the count spectrum $J(q)$ and the electron flux spectrum $F(E)$, viz. $\gamma = \delta + 1$ \citep[Eq.~(1.24) of][]{pianabook}. By comparison, for \emph{thick}-target emission in a target with a (collisional) mean free path that is proportional to $E^2$, the target-averaged electron flux spectrum is correspondingly two powers harder than the \emph{injected} electron spectrum \citep[Eq.~(1.19) of][]{pianabook}, so that $\gamma$ is related to the power-law index $\delta_{\rm inj}$ of the injected spectrum by the oft-cited relation $\gamma = (\delta_{\rm inj} - 2) + 1$, i.e., $\gamma = \delta_{\rm inj} - 1$ \citep[Eq.~(1.25) of][]{pianabook}. 

Table~\ref{tab:spectral-indices} compares the spectral indices provided by spatially integrating the count images, the electron flux spectral images, and the regularized count images with the ones provided by the standard spectroscopy procedure implemented in OSPEX, in the case of five events characterized by similar morphologies. 
As for both the spatially-integrated count images and regularized count images, the corresponding spectral index has been fitted assuming a unit detector response matrix. This approximation does not significantly affect the outcomes of this analysis since, over the bulk of the energy range used to fit the spectral index, the STIX detector response matrix is approximately diagonal with entries that vary over a modest range from \(\sim\)0.36 to \(\sim\)0.42), insufficient to cause the spectral index of the count spectrum to differ significantly from that of the incident photon spectrum. The entries in the table show that 1) the count spectral indices generated by the regularized count images reproduce well the ones provided by OSPEX; and 2) the approximate thin-target relation $\gamma = \delta + 1$ are both well satisfied by the regularized electron flux and count spectra.

\subsection{STIX vs. RHESSI}\label{stix-rhessi}

RHESSI and STIX are both visibility-based instruments, which in general offer the possibility of applying Fourier-based methods for image reconstruction and, more specifically, to use the approach described in Sect.~\ref{S-methodology} to create regularized electron flux and photon spectrum images. However, while the ways RHESSI and STIX generate visibilities both involve occulting grids, they are intrinsically very different. Each of the RHESSI collimators samples a circle in the $(u,v)$-plane via rotation of the spacecraft. In the case of the STIX instrument, each sub-collimator is designed to measure a single Fourier component. As a consequence, while RHESSI in principle can measure hundreds of visibilities, some of them have an insufficient signal-to-noise ratio, and all of them suffer a blurring effect related to the data stacking process; by contrast, STIX provides $30$ complex visibility values that are observed at all times, with better statistics and without any blurring effects.

The higher quality of STIX imaging spectroscopy with respect to RHESSI is demonstrated in Table~\ref{tab:residuals}, where we consider three events observed by STIX and three events observed by RHESSI, all with comparable statistic and morphologies. We reconstructed both the regularized electron flux and regularized count images at different energies using CLEAN \citep{Clean,2023ApJS..268...68P}. For each event and each energy channel we identified the maximum in the residuals map and in the CLEANed map as defined in \cite{2009ApJ...698.2131D} and computed the ratio; then for each event, we averaged this ratio across all energies and computed the corresponding standard deviations. The results in Table~\ref{tab:residuals} show that, for all cases, this ratio is systematically and significantly higher for RHESSI than for STIX, which explains the higher statistical reliability of STIX visibilities when used to obtain the regularized electron spectrum images and count images. 

\begin{table*}[h]
\begin{center}
\begin{tabular}{ccc}
\toprule
Event &  counts & electrons \\
 \midrule
  SOL2022-09-29T14 (STIX) 
& 0.061 $\pm$ 0.007 & 0.069 $\pm$ 0.019 \\
  SOL2002-02-20T11
(RHESSI) & 0.114 $\pm$ 0.027 & 0.171 $\pm$ 0.067 \\
\midrule
 SOL2023-01-11T01
(STIX) & 0.058 $\pm$ 0.012 & 0.065 $\pm$ 0.035 \\
 SOL2003-12-02T22
(RHESSI) & 0.124 $\pm$ 0.025 & 0.292 $\pm$ 0.225 \\
\midrule
 SOL2022-11-11T01
(STIX) & 0.070 $\pm$ 0.013 & 0.084 $\pm$ 0.032 \\
SOL2002-02-20T11
(RHESSI) & 0.128 $\pm$ 0.029 & 0.167 $\pm$ 0.038 \\
\bottomrule
 \end{tabular}
 \caption{Analysis of CLEAN reconstructions for regularized counts and electron flux spectral images obtained using RHESSI and STIX visibilities. The second and third columns show the ratio of the residual and CLEANed maps peaks for the regularized count and electron flux spectral images, respectively, averaged across count and electron energies.}\label{tab:residuals}
\end{center}
\end{table*}

\section{Analysis of the SOL2021-05-08T18 event}\label{S-5-8-21}

For most flares the non-thermal hard X-ray emission from the chromospheric foot-points outshines the much fainter coronal sources at higher energies $\gtrsim 20$~keV, making an analysis of the variation of the electron spectrum along the flare loop very difficult, especially for visibility-based imaging systems such as STIX that necessarily see every source in the field of view. On the other hand, coronal thick-target flares \citep[see][]{2004ApJ...603L.117V} have a higher-than-normal density in the corona, so that very few electrons have sufficient energy to reach the chromospheric footpoints. Consequently, most of the non-thermal emission is from the relatively dense corona, and the resulting absence of footpoint emission makes it feasible to observe how the shape of the (line-of-sight-integrated) nonthermal mean electron flux spectrum $\int F(x,y,z;E) \, dz$ varies along the projection of the flaring loop on the plane of the sky. In such cases the inferred variation of the local electron spectrum $F(E,s)$, at distance $s(x,y)$ from the injection point along the direction defined by the magnetic field lines linking the coronal and chromospheric region of the flare, allows electron transport models to be tested \citep[e.g.,][]{2001ApJ...557..921E,2017ApJ...851...78A}. In the following we show preliminary electron spectra from a flare observed by STIX and characterized by a prominent non-thermal component from the corona, similar to the prototypical coronal thick-target flares from \cite{2004ApJ...603L.117V}. This analysis takes advantage of the enhanced imaging dynamic range provided by regularized imaging spectroscopy.

On May 2021 two active regions generated a series of C- and M-class flares observed by STIX, GOES, and SDO/AIA \citep{2022SoPh..297...93M}. We concentrate here on the event recorded on May~8 in the time interval (relative to Solar Orbiter) between 18:29 and 18:32~UT (SOL2021-05-08T18). The top left panel of Fig.~\ref{fig:may8} shows the (normalized) light-curves for both GOES and two STIX energy ranges, for the time interval considered for this analysis. 
The top right panel shows the AIA image, re-projected to match the viewing angle from Solar Orbiter/STIX and, overlaid, the contours inferred through application of the MEM$\_$GE algorithm to STIX electron visibilities integrated between 25 and 28~keV. Since the flare ribbons are extended, projection effects could have an effect on the interpretation of the actual 3-D structure of the event.  Nevertheless, a loop-like structure connecting the UV flare ribbons, and visible over a wide range of electron energies, is evident in the STIX images.

The bottom left panel of Fig.~\ref{fig:may8} shows the electron flux spectral image in the energy channel between 25 and 28~keV, computed using the regularized imaging spectroscopy code; superimposed are crosses representing eight selected locations within the loop structure. In the bottom middle and right panels, we show the spatially resolved electron flux spectra at each of these locations, with the middle and the right panels showing the spectra corresponding to the locations in the left and right branches of the loop, respectively. The spectra are obtained by averaging the pixel values inside a box of 5 $\times$ 5 pixels around the locations highlighted in the bottom--left panel of Fig.~\ref{fig:may8}. The uncertainties shown by the error bars are obtained by means of the confidence strip approach \citep{1994A&A...288..949P}. The input STIX count visibilities are perturbed 25 times with Gaussian noise with a standard deviation equal to the experimental uncertainty. Then, the electron visibilities and the corresponding electron maps are retrieved from each perturbed data and finally the standard deviation of the spectral intensities obtained from the different electron map realizations is computed. 

\begin{figure*}[t]
    \centering    \includegraphics[width=0.9\textwidth]{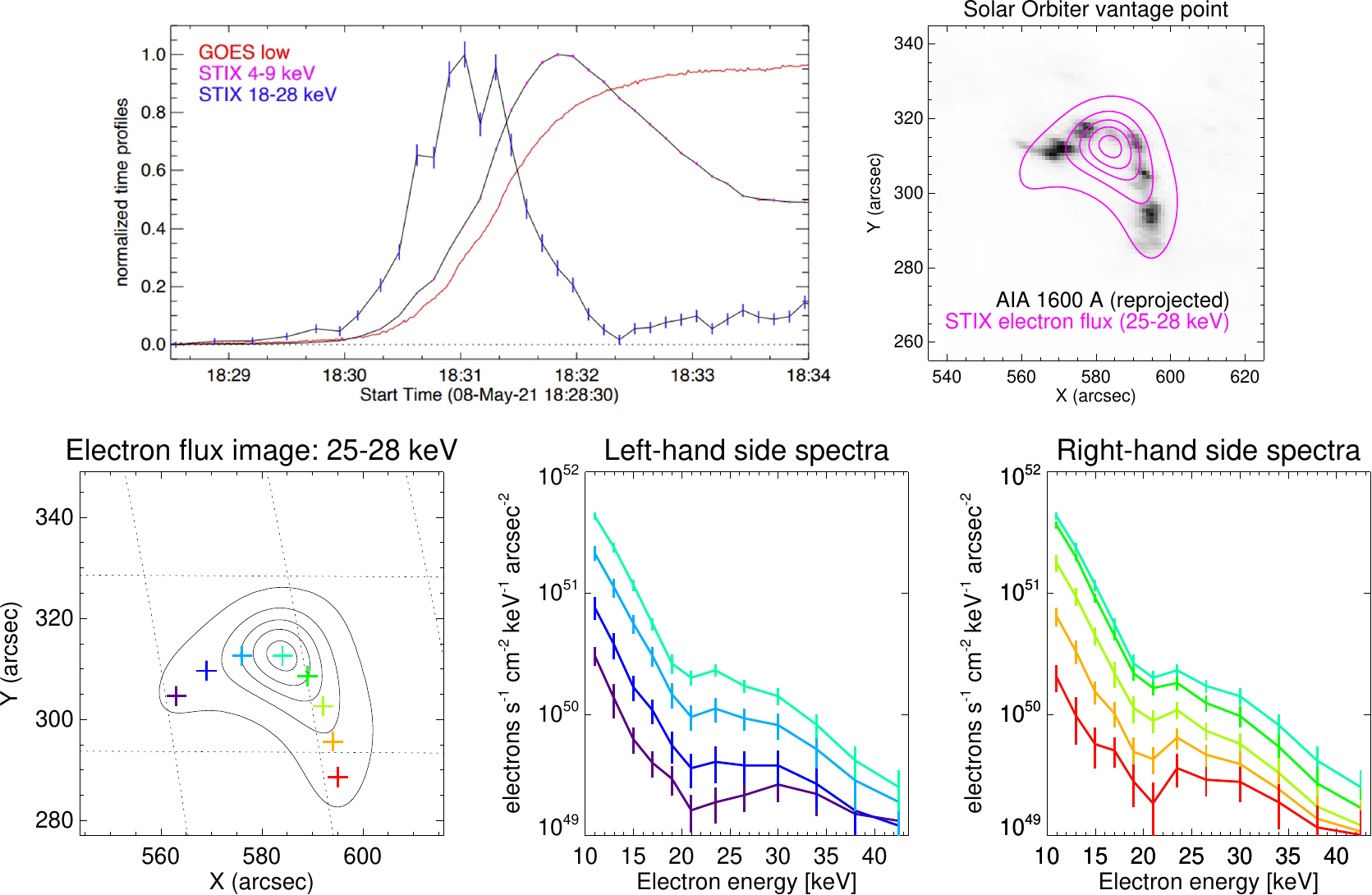}
    \caption{The SOL2021-05-08T18 event.
    \emph{Top left panel}: normalized GOES and STIX time profiles.  \emph{Top right panel}: X-ray contour plots, provided by the MEM$\_$GE algorithm, overlaid on the AIA map, which has been re-projected to match the viewing angle from STIX. The $10$, $30$, $50$, $70$, and $90$\% contour levels of the reconstructed non-thermal X-ray emissions (25 - 28~keV) are plotted in magenta. \emph{Bottom left panel}: electron flux map in the energy channel 25-28~keV, with seven locations identified in the loop. \emph{Bottom middle panel}: regularized electron flux spectra at four locations in the left side of the flaring loop with corresponding uncertainties. \emph{Bottom right panel}: regularized electron flux spectra at five locations in the right side of the flaring loop, with corresponding uncertainties. The spectra are obtained by considering areas of $5 \times 5$ arc seconds around each of the selected points and averaging the corresponding spectra. The spectra clearly show two components: a thermal component below $\sim$20~keV and a non-thermal component at higher energies.
    }
    \label{fig:may8}
\end{figure*}

The spectra clearly show two components: a thermal component below $\sim$20~keV and a non-thermal component above $\sim$20~keV, with a clear minimum, or ``dip,'' in the overall spectrum around 20~keV. It has been shown \citep{2005SoPh..232...63K} that such dips can be an artefact of neglecting to subtract the albedo component \citep[that results from scattering of primary source photons from the photosphere;][]{1978ApJ...219..705B} from the observed hard X-ray spectrum. However, such artefacts are most pronounced in sources with fairly hard ($\gamma \simeq 2$) spectra, which have a proportionately larger number of higher energy photons that backscatter into the energy range being observed; therefore, given the much steeper spectrum ($\gamma \simeq 5.5$; Table~\ref{tab:spectral-indices}) associated with this event, the effects of neglecting albedo photons is unlikely to be the reason for the observed dip. Further, Fig.~4 of \cite{2002SoPh..210..407A} shows that the albedo correction is most pronounced at energies substantially larger than 20~keV; even for flat photon spectra the correction to the inferred electron flux spectrum at 20~keV is only some 10\%, not enough to create an artificial dip of size comparable to those in Fig.~\ref{fig:may8}. We also note that the albedo source typically has a significantly larger spatial extent than the primary source \citep{2002SoPh..210..273S} and therefore is unlikely to make a substantial contribution to the electron flux spectrum in a relatively localized region, such as those considered for computing the spectra shown in Fig.~\ref{fig:may8}.

We therefore conclude that the observed dips are indicative of a real minimum in the primary source electron flux spectrum. Further, the evolving shapes of the nonthermal components show a phenomenological behavior that is plausibly consistent with the effects of electron dynamics within the loop: as the distance from the loop top (the purported location of the electron acceleration region) increases, electrons of progressively higher energies are stopped, slowed, or thermalized through interaction with the electrons in the ambient plasma. Thus the value of the electron flux decreases at all energies, the local minimum at the interface between the thermal and nonthermal components becomes more pronounced, and the energy corresponding to the maximum in the nonthermal component shows a tendency to increase. We shall study this behavior, and its consistency with a collisional model \citep[e.g.,][]{1972SoPh...26..441B,1978ApJ...224..241E} for the electron dynamics, more quantitatively in a subsequent work. 

\section{Comments and Conclusions}\label{S-conclusions}

The first objective of this study was to show that the regularized imaging spectroscopy approach introduced for the analysis of RHESSI visibilities is also feasible in the STIX framework. Further, we have shown that the peculiar one-to-one correspondence between visibilities and collimators realized by the Moir\'e pattern technique utilized in STIX allows the generation of research products (namely, regularized electron spectral images and count images, and regularized spatially resolved electron flux and count spectra) that have significantly better quality than those obtained with the rotating modulation collimator approach of RHESSI. Thanks to this higher reliability, we have been able to show, for the first time, how the spectrum of the accelerated electron flux varies along the path from the corona to the dense chromospheric regions at the loop footpoints, presumably through interaction of the accelerated electrons with the ambient plasma.

One aspect that may hamper the reliability of the outcomes of this approach is the limited dynamic range provided by the STIX detectors. Indeed, the spectra in Fig.~\ref{fig:may8} show ranges of intensities which, for all energy channels, are almost always close to this 10:1 limit, across all locations selected within the loop. 
Nevertheless, it is clear that this sort of analysis cannot be performed for all flares, since frequently the dynamic range is a limiting factor in constructing accurate spectra. For example, such an analysis will be likely impossible to perform for flares with prominent footpoint emission, since footpoint emission is typically much brighter than the emission from the flare legs. The best candidates would be flares with a prominent coronal source such as coronal thick-target sources \citep{2004ApJ...603L.117V} or flares with occulted footpoints. For the latter a case, a second observatory that sees the entire flare site \citep[see, e.g.,][]{2019RAA....19..167K} could provide additional information on the spectral shape of the footpoints.

Regularized imaging spectroscopy with STIX visibilities is therefore probably the first\footnote{An even more reliable analysis can be performed using observations that will be provided by the Focusing Optics X-ray Solar Imager (FOXSI) \citep{2023BAAS...55c.065C}, a direct imaging telescope that is characterized by a potentially unprecedented sensitivity for hard X-rays.} technique that could allow the testing of energy loss rates \citep[for example, either Coulomb or non-Coulomb collisional losses; see Sect.~4 of][]{2001ApJ...557..921E} for non-thermal electron transport. Given that such loss rates in general depend on the ambient density, future work will be devoted to investigate whether it is possible to determine the ambient density through the dependence of the electron spectrum on position, and thus separate the mean electron flux and the ambient electron density factors in the bremsstrahlung equation~(\ref{eq:fundamental}). The thus-determined ratio of the accelerated electron population to the ambient density in the flaring region constitutes an important constraint on the effectiveness of electron acceleration mechanism(s) in solar flares.

\section*{Acknowledgements}
\label{sec:acknowledge}
{\em{Solar Orbiter}} is a space mission of international collaboration between ESA and NASA, operated by ESA. The STIX instrument is an international collaboration between Switzerland, Poland, France, Czech Republic, Germany, Austria, Ireland, and Italy. SK is supported by the Swiss National Science Foundation Grant 200021L\_189180 and the grant ``Activit\'es Nationales Compl\,ementaires dans le domaine spatial'' REF-1131-61001 for STIX. 
PM and AGE are supported by NASA award number 80NSSC23M0074, the NASA Kentucky EPSCoR Program and the Kentucky Cabinet for Economic Development; AGE also acknowledges support from the NASA Heliophysics Supporting Research Program under award 80NSSC23K0448. AV, MP, and AMM acknowledge the support of the “Accordo ASI/INAF Solar Orbiter: Supporto scientifico per la realizzazione degli strumenti Metis, SWA/DPU e STIX nelle Fasi D-E”. The authors thank Isaiah Beauchamp and Jana Kašparová for several useful discussions.

\bibliographystyle{aa}
\bibliography{bib_stix}

\end{document}